\documentclass[conference]{IEEEtran}
\IEEEoverridecommandlockouts
\usepackage{cite}
\usepackage{amsmath,amssymb,amsfonts}
\usepackage{algorithmic}
\usepackage{graphicx}
\usepackage{textcomp}
\usepackage{float}
\usepackage{xcolor}
\def\BibTeX{{\rm B\kern-.05em{\sc i\kern-.025em b}\kern-.08em
    T\kern-.1667em\lower.7ex\hbox{E}\kern-.125emX}}

\title{Effects of Distributed Friction Actuation\\ During Sliding Touch}

\author{MacKenzie Harnett,
        Paras Kumar,
        Rebecca F. Friesen*,~\IEEEmembership{Member,~IEEE.}
\thanks{ Authors are with the Department of Mechanical Engineering, Texas A\&M University, USA      (e-mail: rfriesen@tamu.edu). }
\thanks{* Corresponding author.}
\thanks{Manuscript received XXXXX; revised XXXXXXX.}}

\begin{document}
\maketitle
\begin{abstract}
Friction modulation allows for a range of different sensations and textures to be simulated on flat touchscreens, yet is largely unable to render fundamental tactile interactions such as path following or shape discrimination due to lack of spatial force distribution across the fingerpad. In order to expand the range of sensations rendered via friction modulation, in this paper we explore the possibility of applying spatial feedback on the fingerpad via differing friction forces on flat touchscreens. To this end, we fabricated six distinct flat surfaces with different spatial distributions of friction and observed deformation of the fingerpad skin in response to motion along these physical samples. In our study, friction changes that occur sequentially along the sliding direction introduced little transitory spatial warping such as compression or stretching to the fingerpad, suggesting limited perceptual differences in comparison to 'classic' friction modulation. Distributing friction across the direction of motion, however, showed pattern-dependent shearing of the fingertip skin, opening avenues for new sensations and illusions heretofore unachievable on flat touchscreen surfaces. 
\end{abstract}

\begin{IEEEkeywords}
Force Rendering, Friction Modulation, Surface Haptics
\end{IEEEkeywords}

\section{Introduction}
In the modern era, touchscreen devices are ubiquitous and essential for many day-to-day tasks; touchscreens are on our phones and tablets, in elevators and cars, and anywhere else we might want to push a button or read a display. They make information and education more accessible and support unprecedented levels of socialization, global engagement, and communication. However, this extensive integration and dependency on touchscreens has led to a sharp decline in haptic feedback often associated with more analog devices. This feedback is necessary for dexterous movements, natural feeling sensations, and complex touch-based tasks. These are all essential components of interpreting the world around us; thus, it is highly beneficial to develop a mechanism by which this feedback can be reproduced on touchscreens. 

A promising method of applying haptic actuation on flat screens is friction modulation. Many recent friction-modulated screens exploit an electroadhesive effect to increase friction forces on a bare finger as a function of applied voltage~\cite{bau2010teslatouch, vardar2017effect, shultz2018application}. This effect has also been accomplished by dynamically reducing friction forces via ultrasonic vibration of the screen~\cite{watanabe1995method, winfield2007t, wiertlewski2016partial}.
Regardless of the specific approach, friction modulation has the advantage of applying strong, high bandwidth lateral forces directly to the fingertip when in contact with a screen. Significant advantages include circumventing  the need to use an intermediary tool---such as a stylus or wearable device~\cite{culbertson2014modeling, friesen2023perceived} and more naturalistic sensations relative to vibrotactile feedback, rendered directly at the fingertip~\cite{choi2012vibrotactile}.

\begin{figure}[ht]
\centerline{\includegraphics{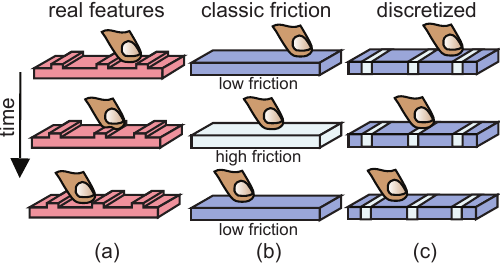}}
\caption{\textbf{Different Friction Rendering Strategies.} Illustration comparing real textured surfaces, classic friction modulation, and discretization of a friction modulating surface.  (a) a finger slides over a real surface with physical ridges, (b) a finger slides over a friction modulating screen simulating the leftmost real surface by applying a 'High' or 'Low' friction force onto the fingerpad, (c) a finger slides over a friction modulating screen, which is segmented into discrete sections to simulate the sensation of the fingerpad as it is sliding on the border between two distinct friction states.}
\label{frictioncomp}
\end{figure}

Despite their many capabilities, modern friction modulating screens are limited in the range of sensations that they can generate \cite{friesen2021building}, including an inability to effectively facilitate fundamental tasks such as shape recognition, path following, or edge detection \cite{gershon2016visual, Sadia2022, Osgouei2016}. Such interactions may require a spatial distribution of forces across the fingerpad. While classical implementations of friction modulation can have high spatio-temporal resolution with respect to finger position, they apply only one friction state to the entire fingerpad at any given time, even when attempting to render inter-fingerpad differences as depicted in Figure~\ref{frictioncomp} (a)-(b). 
Virtual texture rendering is limited by current friction modulation techniques as well; several research groups have proposed that the limitations of current generation (hereafter referred to as 'classic') friction modulating screens in generating virtual textures are due to a lack of spatio-temporal feedback within the fingerpad \cite{friesen2021building,wiertlewski2011spatial}.
The integration of spatial feedback within the finger contact patch has the potential to improve rendering of existing textures and produce a broader library of potential sensations on touchscreens. 

In recent years, a variety of methods for providing spatially distributed feedback have been proposed. While previous work has documented some perceptual effects of spatially distributed changes in oscillating friction force on the bare fingerpad via a flat touchscreen~\cite{Hiroki_Ishizuka2017}, none, to the best of our knowledge, have characterized the \textit{physical} response of the fingerpad to static distributions of friction force.
One group has observed physical displacement of friction-actuated pucks worn between a finger and a screen~\cite{tan2021soft},\cite{tan2023pixelite}. Their electroadhesive pucks actuate different sections of the fingerpad skin, and they found that adjacent sections of the fingerpad would displace along with the one under the actuated puck, resulting in participants struggling to discriminate between neighboring pucks. 
Both ultrasonic and electroadhesive friction modulating screens with segmented friction feedback have been developed ~\cite{ilkhani2018creating, katumu2016using}, but only with coarse discretization for multi-touch applications.

In order to thoroughly evaluate the potential effectiveness of high resolution multi-electrode friction modulating screens, we constructed discretized, multi-friction surfaces and examined the physical response of the fingerpad when travelling along these surfaces. Figure~\ref{frictioncomp} (c) conceptualizes an alternative approach to single-electrode friction modulating surface, instead using spatially distributed friction patterns. 
We anticipate that the outcome of this work will inform future design of friction modulating screens, in the event that discretized friction surfaces expand and enhance the user tactile experience. 
The work presented here utilizes samples with real, physical changes in friction; we opted to first observe fingerpad behavior on real friction patterns as opposed to immediately developing multi-electrode friction modulating screens. We have done this to verify the presence of fingerpad behaviors unique to discretized friction surfaces so as to test the merit of the proposed touchscreen design prior to manufacturing. Through this we aim to explore the potential of 2D tactile screens in simulating a broader range of sensations in the tactile space and facilitate novel 2D virtual tactile environments, such as rendering distinct shapes or enabling edge detection. 

\section{Methods and Materials}

When building test surfaces that have discrete areas with distinct friction properties, we set two primary design objectives. Firstly, we designed surfaces to be as transparent as possible so as to enable visual capture of skin deformation within a finger contact patch. Secondly, we looked for surface finishes with substantially different friction coefficients. 
Using a finely textured surface coating to lower friction, we generated six samples with distinct interface geometries to determine whether and how the fingerpad contact area deforms when straddling multiple friction states. 

\subsection{Sample Preparation}
\begin{figure}[ht]
\centering
\includegraphics{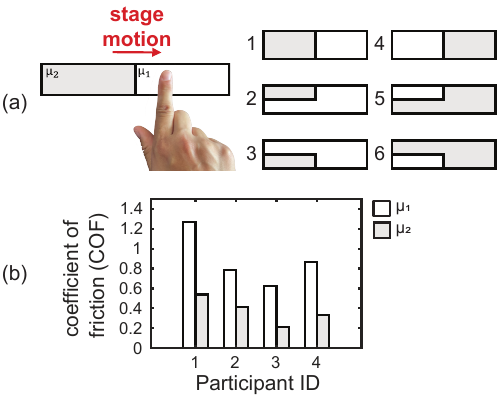}
\caption{\textbf{Sample Configurations} (a) Geometries and distribution of friction for each of the six samples evaluated in this study; (b) Coefficient of Friction (COF) measurements for all four participants. Average COF for the high and low friction samples was calculated to be $\mu_1=0.886$ and $\mu_2=0.376$, respectively.}
\label{fig_samples}
\end{figure}

A total of six 25x150\,mm$^2$ surface geometries were constructed and are illustrated in Figure~\ref{fig_samples}. Each sample consists of acrylic pieces with two different friction coefficients. Lower friction pieces were treated with a clear spray with a satin finish (Krylon K05562007, Satin) to lower the coefficient of friction (COF) while maintaining sample transparency, and all pieces were leveled to the same height. The first three samples apply an overall decrease in friction at the transition, while the remaining three apply an increase in friction. Two of the samples (1, 4) orient the only resultant change in friction sequentially along the direction of motion, and the remaining four (2, 3, 5, \& 6) also include a change in friction across the direction of motion. 

A tribometer, which consisted of a lateral stage (Thorlabs DDS300) and two force sensors (ATI nano43), was used to measure the coefficient of friction (COF) for both the treated and untreated acrylic surfaces across all study participants. The surfaces were moved laterally under each fingerpad at 20\,mm/s with a normal force between 0.2-0.3\,N, mirroring the conditions of the experimental procedure. A light dusting of cornstarch powder was applied to the fingerpad prior to these measurements, again to mirror experimental conditions. The lateral and normal force data was collected, averaged across 0.7 seconds of data (1.4kHz sampling rate), and the ratio of the lateral and normal forces was used to calculate the COFs. These values were found to be, on average, 0.886 for the plain acrylic sample and 0.376 for the treated sample across participants (see: Figure~\ref{fig_samples}). 

\subsection{Marker Application}
\begin{figure}[t]
\centering
\includegraphics{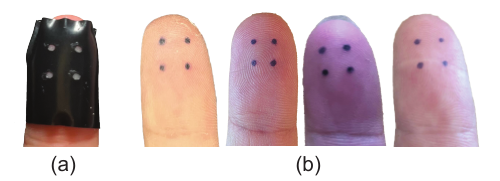}
\caption{\textbf{Marker Application} (a) finger with stencil, (b) participant fingers post-tattoo application. }
\label{fig_mark}
\end{figure}

Prior work~\cite{huloux2021measure,barrea2018perception,tada2004imaging,delhaye2016surface} has implemented an 'optical flow' method to evaluate fingerpad deformation through the motion of landmarks on the fingerpad (e.g., fingerprint ridges). While this technique yields high resolution of strain, it was impractical in our case due to limited visibility through our samples caused by our friction-lowering textured spray. Instead, we opted for dark markers applied to the fingerpad to ensure uninterrupted tracking of four distinct points through the semi-transparent samples \cite{kaneko2020measurement}.
A 2x2 marker grid was applied to the tip of the right index finger using temporary tattoo ink (INKBOX). Each marker has a diameter of 1.5\,mm and a center-to-center spacing of 4\,mm. A custom stencil was used to ensure precise application of the marker grid as shown in Figure~\ref{fig_mark}. Unlike traditional ink, the temporary tattoo ink penetrates the epidermis and does not smear during sliding touch. 

\section{Data Collection}\label{section:expsetup}

\begin{figure}[t]
\centering
\includegraphics{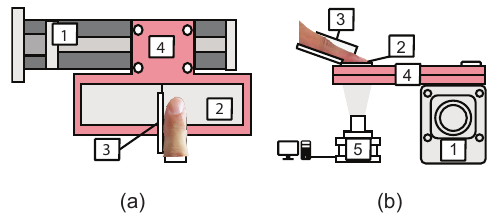}
\caption{\textbf{Experimental Setup} Illustration of final experimental setup. (a) Top view, (b) Side View. (1) Motorized Linear Stage, (2) Sample, (3) Finger Brace, (4) Sample Reservoir/Mount, (5) Camera. We only record data as the reservoir moves to the right, relative to the reader.}
\label{fig_setup}
\end{figure}
We recorded and evaluated the deformation of the marker grid tattooed on the fingerpad for the six different sample configurations across four participants (2 female, ages 24-37). 
This study was approved by the Institutional Review Board of Texas A\&M University (IRB2024-0539), and participants were compensated for their participation. 
We began with a small number of participants due to the time intensive nature of the study (approximately four hours), multi-day tattoo application procedure, and because the study did not collect subjective judgements prone to additional noise. 

A minimum of one day before testing, each participant's right index finger was tattooed with the 2x2 marker grid. Each participants tattooed fingerpad is presented in Figure~\ref{fig_mark}. Immediately before testing, participants were asked to practice applying a normal pressing force kept within the range 0.2---0.3\,N to a flat surface affixed to a force sensor (SingleTact 8\,mm Diameter, 10\,N/2.2\,lb Force). The applied force was monitored on an external display. After participants were able to consistently apply a force within the specified range a minimum of three times in a row, they proceeded to data collection. 

Prior to interacting with any sample, participants were asked to wash their hands with isopropyl alcohol and apply a light dusting of cornstarch-based powder (Clubman) to their fingerpad to reduce sweat buildup and stick-slip effects. Each surface sample was secured within the frame of a moving platform to ensure constant height, then moved laterally with respect to a stationary finger via a motorized stage. Index fingers were braced against a finger holder to ensure the finger up to the distal joint remained fixed, and the participant focused on maintaining constant pressing force throughout motorized swipes. Samples were actuated along the participant's fingerpad at a speed of 20\,mm/s for a total distance of 100\,mm. For the duration of each trial, fingerpad deformation was recorded using a camera (Imaging Source DFK 37BUX273) mounted directly below the sample. The camera had a maximum resolution of 1440 x 1080 pixels (px) and a frame rate of 30 frames-per-second (FPS). After three trials, the next sample configuration was placed in the sample reservoir and the procedure was repeated. A total of 18 trials were collected and evaluated for each participant. The experimental setup is illustrated in Figure~\ref{fig_setup}.

\subsection{Image Processing and Marker Tracking}

\begin{figure}[t]
\centering
\includegraphics{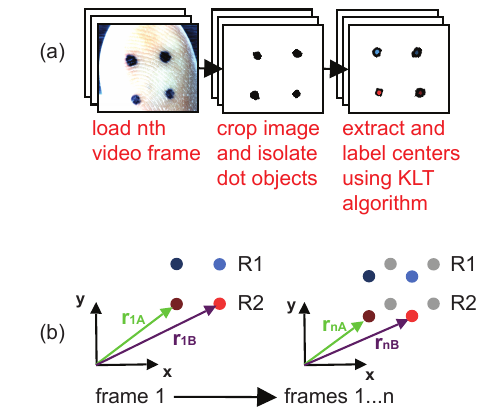}
\caption{\textbf{Marker Tracking} (a) Visualization for Image Processing and labelling of 2-by-2 dot grid. (b) Visualization of total dot displacement values used in absolute and relative displacement calculations}
\label{fig_track}
\end{figure}

Videos were cropped to the area of interest. Each frame was thresholded to convert into a black and white image, isolating the four markers from surrounding noise. These markers were labelled, sorted, and their positions tracked using the Kanade-Lucas-Tomasi (KLT) feature-tracking algorithm ~\cite{lucas1981iterative, tomasi1991detection} in Matlab. This process is illustrated in Figure~\ref{fig_track}(a). 
The initial position of each marker centroid was extracted from the first frame of the video and used to determine the pixel to mm scaling ratio using the 4mm absolute distance between markers. For each frame, the marker centroids were identified, demarked with a colored dot to indicate accurate tracking, and their (x, y) values stored in an array. 

Data was low-pass filtered ($f_{pass}=5\,Hz$) to remove mechanical noise. The frame at which the leading edge of dots crossed the friction transition was manually extracted from the video data and labeled as time t1. The trailing edge of dots hits the friction transition 5 frames later (0.167 seconds) at time t2. We refer to the time between t1 and t2 as the transition period, during which the fingerpad is spanning the friction transition. 

Dot displacement for the nth frame, $d_{n}$, was calculated as the difference between current position $r_{n}$ and the position at frame one, $r_{1}$. 
\begin{equation}
    d_{n}= r_{n}-r_{1} 
\end{equation}
\noindent Here, frame one is defined as five frames before the friction transition hits the leading edge of the finger, at which point the skin had reached a steady state sliding on the first friction region.
Relative displacement is calculated as the difference between two positions A and B using the corresponding variables defined in Figure~\ref{fig_track}(b): 
\begin{equation}
    d_{nAB}= r_{nA}-r_{nB}   
\end{equation}
\noindent When calculating final overall displacement, we compared the average displacements from the five frames immediately preceding and following the transition period:
\begin{equation}
    \Delta d_{final}= \frac{\sum_{m}^{m+4} d_{i}}{5}-\frac{\sum_{1}^5 d_{i}}{5}   \label{eq:3}
\end{equation}
\noindent Here, m corresponds to the frame at which the trailing edge of markers crosses the transition region. 

\section{Results and Discussion}

\begin{figure}[t]
\centering
\includegraphics{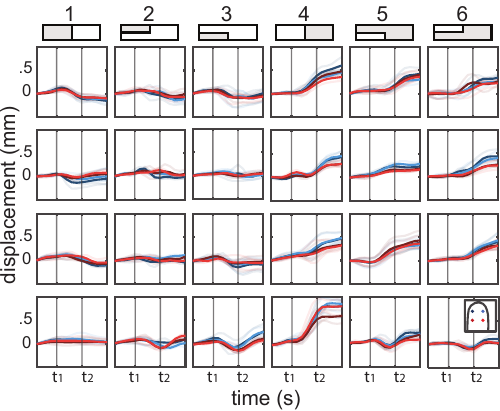}
\caption{\textbf{Total Displacement Plots} Total displacement plots summarizing all participants, samples, and trials. Each row of plots corresponds to the averaged results across three trials for each participant, with the results of each individual trial in light grey. The vertical grey lines indicate when the leading and trailing columns of dots hits the friction transition, respectively. For all plots, t2-t1 = .167 seconds.}
\label{fig_cons}
\end{figure}

Dot displacement for all participants, samples, and trials is summarized in Figure~\ref{fig_cons}.
Skin behavior for a given participant across trials was consistent; the similar behavior of individual trials are shown in lighter gray and averaged for each participant and sample combination. Behavior across participants was more variable despite efforts to ensure consistent touch conditions. In the following analysis, we observe trends in both transitory and sustained skin deformation across different sample types. 

\subsection{Transitory Friction Change}\label{section:absMarkerDisp}

\begin{figure}[t]
\centering
\includegraphics{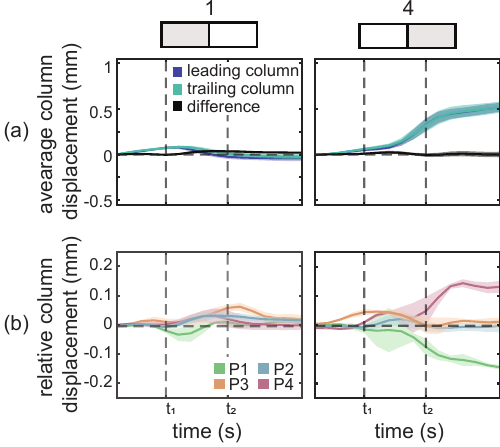}
\caption{\textbf{Relative Displacement Plots} Relative displacement plots for samples 1 and 4. (a) Comparison of relative displacement with total column displacement, averaged across all participants. (b) Zoomed-in comparison of participant results for relative displacement. Negative displacement indicate compression of the marker grid, positive displacement indicates stretching of the grid. Shaded region represents the standard error.}
\label{fig_rel}
\end{figure}

When sliding across samples 1 and 4, the finger experiences a transitory period with multiple distributed friction forces before returning to a single, new friction force. 
Across all samples, the grid, on average, moved in bulk as the leading edge of the fingerpad transitioned onto the succeeding friction state. That is, all dots began displacing at the same point in time, regardless of when they individually hit the transition. This is summarized in Figure~\ref{fig_rel}(a), which shows that difference between the leading and trailing columns varied very little over the transition region, especially relative to total displacement. 

The magnitude of change in total displacement was more dramatic for low-to-high transitions than for high-to-low transitions.
We did see differences in skin behavior between individual participants; Figure~\ref{fig_rel}(b) shows that participants 1 and 4 had compression or stretching, respectively, between the columns as the trailing edge crossed the transition period. Regardless of these differences, the largest changes in relative column displacement are sustained post-transition, indicating that they are due not to the distributed friction states but to the overall change in friction. 

These results demonstrate that sequential changes in friction along the fingerpad's motion induce primarily simultaneous bulk movement of the fingerpad, 
which would not deviate significantly from the behavior expected on a single electrode friction modulating screen.
Therefore, we anticipate limited benefits from distributing friction along the fingerpad path for the purposes of new sensations enabling edge detection or contour rendering. 
Additionally, the magnitude of total displacement for high-to-low changes in friction was much less dramatic than seen in the low-to-high cases despite an equal absolute change in friction. Deformation behavior is thus dependent on the order in which different friction states are encountered. 

\subsection{Sustained Friction Change}

\begin{figure}[t]
\centering
\includegraphics{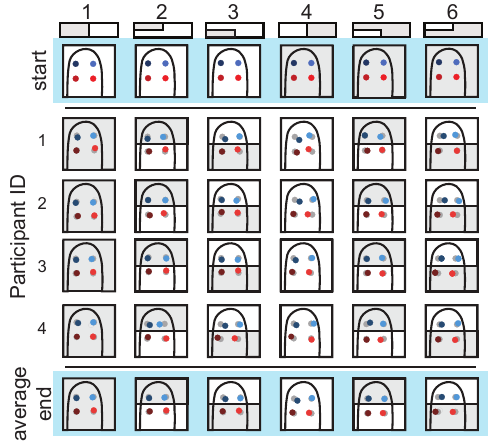}
\caption{\textbf{Deformation Visualization}. Visualization of the final deformation of each dot grid for each sample. The middle four rows depict final relative deformation for each participant, averaged across 3 trials. The final row represents the averaged results across participants. To improve visibility, displacements were scaled by a factor of 10.}
\label{fig_defvis}
\end{figure}

To more thoroughly understand sustained changes in deformation, we 
visualized final marker displacement relative to starting position; see equation \ref{eq:3}.

These averaged relative displacements were then used to recreate the final deformed state of the fingerpad, which is illustrated in Figure~\ref{fig_defvis}. 
As we anticipated, there was minimal movement along the y-axis, perpendicular to swipe direction. 
 
Instead, much of the resultant deformation manifested as shearing between the top and bottom rows.

Changes were particularly dramatic for the low-to-high changes, where shearing was readily observed across all samples. This change was less dramatic when only the bottom row of dots cross onto a high friction state (sample (5)), as opposed to greater shearing when only the top row hits a high friction state (sample (6)). 

\begin{figure}[t]
\centering
\includegraphics{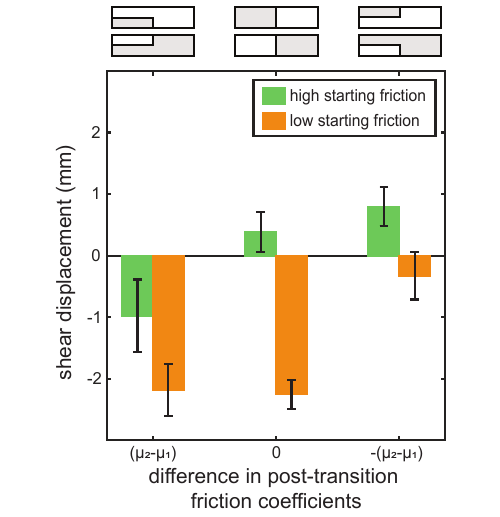}
\caption{\textbf{Shear Barcharts}. Results for shear displacement across 3 trials for 4 participants. The samples corresponding to each COF are depicted above the plot. Error bars correspond to the standard error of each distribution.}
\label{fig_defbars}
\end{figure}

After confirming the existence of shearing with our visualization, we quantified how shearing was affected by different distributions of friction. Shear displacement ($\Delta x$) was calculated as the x-axis difference in displacement between the centers of each row of markers before and after they cross over the sample transition region. 
\begin{equation}
    \Delta x_{avg} = \Delta x_{R1}-\Delta x_{R2}
\end{equation} 

\noindent The difference in the post-transition friction coefficient ($\Delta \mu$) was calculated as the difference between the bottom and top halves of the second half of the sample's friction coefficients. 
\begin{equation}
    \Delta \mu = \mu_{B}-\mu_{T}
\end{equation} 

Summarized shearing is shown in Figure~\ref{fig_defbars}. 
For both the low-to-high and high-to-low cases, the shear displacement increases as the difference in the post-transition friction coefficients becomes more positive. 
Regardless of starting friction, the fingerpad shears strongly in one direction when friction is high on top and low on the bottom row, and shears weakly in the other direction when friction is low on top and high on the bottom row. 
By distributing friction across the fingerpad's path, as was done in samples (2-3) and (5-6), we can influence the resultant shearing behavior of the fingerpad. 

Consistent with the results from ~\ref{section:absMarkerDisp}, the magnitude of shear displacement for the low-to-high samples is generally larger than that of the high-to-low samples, but both show the same trend. 

These results promise new patterns of deformation actuated by statically distributed friction across the fingerpad path. 
 
The ability to control the deformation of the fingerpad can facilitate novel sensations that classic friction modulating screens are incapable of generating due to innate 'whole-finger' changes in friction.

\subsection{Study Limitations and Future Work}

This work focused on fingerpad deformation on \textit{real} textured samples, complicating direct perceptual comparison to temporal changes in whole-finger friction only possible on friction modulating devices. 
Our results motivate construction of spatially-distributed friction modulation surfaces, but future work must validate whether such spatial distributions can create perceptually distinct sensations that mirror distinct patterns in skin deformation.
Furthermore, our experimental setup was limited in low spatial resolution of the 2x2marker grid and low temporal resolution of the 30 fps collection rate; we believe this resolution was adequate for observing gross trends, but subsequent work can more closely examine subtle deformation effects occurring in the short transition window for abrupt friction changes. Increasing grid resolution would allow us to better characterize distributed deformation across the fingerpad under different conditions. 
Finally, this work only observed contact during lateral finger motion in one direction, and for only two friction forces. In vivo interactions with touchscreen surfaces involve sliding the finger in all directions, and it would be interesting to see how shearing occurs along different axes of finger motion and greater ranges of friction differences. 

\section{Conclusion}
This study examined the effects of discrete friction forces actuated across the fingerpad, in order to ascertain the potential of multi-electrode friction modulating screens to manipulate finger skin deformation patterns in new ways. 
We found that discretized friction may be unable to provide the high resolution feedback necessary for edge detection, as the finger skin moves largely in bulk when passing over friction transitions.
There may be opportunities for other novel sensations, however; we observed controllable shearing direction when the finger spanned discrete friction regions distributed perpendicular to direction of motion.
Future work aims to determine the extent by which the fingerpad can be manipulated and deformed, how individuals perceive these distributed forces, and further define potential applications. The imaging method defined in this paper also allows us to image through semi-transparent textured surfaces, which will help characterize and catalog the physical response of the fingerpad in contact with a variety of textured surfaces. 

\bibliography{bibliography}
\bibliographystyle{ieeetr}
\end{document}